\documentclass[journal=nalefd,manuscript=letter]{achemso}

\usepackage[T1]{fontenc}
\usepackage[utf8]{inputenc}
\usepackage[english]{babel}
\usepackage{subfiles}

\usepackage{amsmath}
\usepackage{caption}
\usepackage{subcaption}
\usepackage{hyperref}
\usepackage{orcidlink}
\usepackage{physics}
\makeatletter
\renewcommand*{\acs@author@fnsymbol}[1]{%
  \ensuremath{%
    \ifcase#1
      *%
    \or 1%
    \or 2%
    \or 3%
    \or 4%
    \or 5%
    \or \dagger%
    \else #1%
    \fi
  }%
}
\makeatother

\title{Nanoscale Imaging of Strain-Controlled Altermagnetic Domains in $\alpha$-MnTe}

\author{Alex L.~Melendez\,\orcidlink{0009-0003-1610-1340}}
\affiliation{%
Center for Nanophase Materials Sciences,
Oak Ridge National Laboratory,
Oak Ridge, TN 37831, USA
}
\altaffiliation{These authors contributed equally to this work.}

\author{Sijie Xu\,\orcidlink{0000-0002-5219-9862}}
\affiliation{%
Department of Physics \& Astronomy,
Rice University,
Houston, TX 77005, USA
}
\alsoaffiliation{%
Rice Laboratory for Emergent Magnetic Materials and
Smalley-Curl Institute,
Rice University,
Houston, TX 77005, USA
}
\altaffiliation{These authors contributed equally to this work.}

\author{Liangbo Liang\,\orcidlink{0000-0003-1199-0049}}
\affiliation{%
Center for Nanophase Materials Sciences,
Oak Ridge National Laboratory,
Oak Ridge, TN 37831, USA
}

\author{Rong-Zhu Lin\,\orcidlink{0000-0003-1456-2656}}
\affiliation{%
Department of Physics \& Astronomy,
Rice University,
Houston, TX 77005, USA
}
\alsoaffiliation{%
Rice Laboratory for Emergent Magnetic Materials and
Smalley-Curl Institute,
Rice University,
Houston, TX 77005, USA
}

\author{An-Ping Li\,\orcidlink{0000-0003-4400-7493}}
\affiliation{%
Center for Nanophase Materials Sciences,
Oak Ridge National Laboratory,
Oak Ridge, TN 37831, USA
}

\author{Pengcheng Dai\,\orcidlink{0000-0002-6088-3170}}
\affiliation{%
Department of Physics \& Astronomy,
Rice University,
Houston, TX 77005, USA
}
\alsoaffiliation{%
Rice Laboratory for Emergent Magnetic Materials and
Smalley-Curl Institute,
Rice University,
Houston, TX 77005, USA
}

\author{Hu Miao\,\orcidlink{0000-0003-1078-5713}}
\affiliation{%
Materials Science and Technology Division,
Oak Ridge National Laboratory,
Oak Ridge, TN 37831, USA
}
\email{miaoh@ornl.gov}

\author{Zhaoyu Liu\,\orcidlink{0000-0002-9894-9622}}
\affiliation{%
Department of Physics \& Astronomy,
Rice University,
Houston, TX 77005, USA
}
\alsoaffiliation{%
Rice Laboratory for Emergent Magnetic Materials and
Smalley-Curl Institute,
Rice University,
Houston, TX 77005, USA
}
\alsoaffiliation{%
Department of Chemistry,
Clemson University,
Clemson, SC 29634, USA
}
\email{liuzy.phy@gmail.com}

\author{Huan Zhao\,\orcidlink{0000-0002-4982-0865}}
\affiliation{%
Center for Nanophase Materials Sciences,
Oak Ridge National Laboratory,
Oak Ridge, TN 37831, USA
}
\email{zhaoh1@ornl.gov}

\begin{document}

\begin{abstract}
\textbf{
Altermagnets combine compensated magnetic order with momentum-dependent spin splitting, offering a route to spintronic functionality without the stray fields of conventional ferromagnets. Mechanical strain provides a promising means of controlling their N\'eel order, yet the microscopic pathway by which strain reorganizes an altermagnetic texture remains unresolved. Here, we combine scanning nitrogen-vacancy magnetometry with a piezo-driven uniaxial strain cell to directly image the strain-driven evolution of magnetic domains in bulk $\alpha$-MnTe at room temperature. By applying uniaxial stress along the nearest-neighbor Mn-Mn bond direction, we find that compressive strain reorganizes the magnetic texture through domain coalescence, increasing the size of the largest connected domain while reducing the domain-wall density. On sweeping toward tensile strain direction, however, the domain network follows a distinct trajectory from that observed during the compressive sweep. Instead, the large connected regions fragment into a new metastable configuration, producing pronounced hysteresis in the maximum domain size and stray-field distribution. These results identify domain connectivity and topology as key carriers of strain-induced magnetic memory. Our work reveals domain coalescence and hysteretic fragmentation as the microscopic pathway of strain control in $\alpha$-MnTe and establishes a route toward strain-programmable altermagnetic textures and reconfigurable spintronic devices.
}
\end{abstract}

\maketitle

\section{Introduction}

Altermagnets are a recently identified class of magnetic materials that combine the vanishing net magnetization of antiferromagnets with the spin-polarized electronic structure characteristic of ferromagnets.
\cite{
    Smejkal2022PRX_beyond,
    Smejkal2022PRX_emerging,
    Smejkal2022NatRev,
    songAltermagnetsNewClass2025,
    jungwirth2026symmetry}.
These properties make altermagnets promising candidates for dense, fast and robust spintronic technologies~\cite{RevModPhys.90.015005}.
Among the proposed altermagnetic materials, the $g$-wave altermagnet $\alpha$-MnTe has emerged as a prototypical platform, combining collinear compensated magnetic order above room temperature with momentum-dependent spin splitting and an anomalous Hall effect (AHE) despite its nearly vanishing net magnetic moment along the $c$ axis
\cite{
    Krempasky2024Nature,Mcclarty2024,
    Gonzalez2023PRL,Kluczyk2024PRB,Liu2025arXiv}.

A central challenge for altermagnetic spintronics is the deterministic manipulation of the magnetic order.
Strain engineering of altermagnet (AM) using a voltage-controlled piezoelectric strain cell\cite{hicksPiezoelectricbasedApparatusStrain2014a} has attracted significant interest because magnetoelastic coupling enables dynamical modification of magnetic anisotropy without requiring large applied magnetic fields \cite{Baral2023adfm}.
Recent transport and neutron scattering have shown that uniaxial strain can modify the AHE of $\alpha$-MnTe and lift the degeneracy between crystallographically related magnetic variants
\cite{
    Liu2025arXiv,smolenski2025strain,
    Aoyama2024PRMaterials}.
These observations have been interpreted as strain-induced detwinning: compressive strain generated by uniaxial stress applied along the nearest-neighbor (NN) Mn-Mn bond direction preferentially populates one of the three in-plane N\'eel axes, while a $c$-axis magnetic field selects between the remaining time-reversed variants, driving the system toward a single-domain state in bulk sample \cite{Liu2025arXiv}. 
Measurements in altermagnet candidate h-FeS show similar behavior, i.e, uniaxial compressive strain drives the system into a single domain state in bulk sample~\cite{Yao2026hFeS}.
By contrast, magneto-optical measurements suggest that the twofold magnetoelastic anisotropy can overcome the intrinsic sixfold magnetocrystalline anisotropy and continuously rotate the local N\'eel vector ($\mathbf{L}$) rather than discrete detwinning
\cite{liebman2026strain}.
%
These pictures emphasize different observables: the detwinning picture describes strain-induced changes in domain populations, whereas the continuous-rotation picture describes changes in the local N\'eel-vector angle within the optically probed region.
%
However, neither approach directly resolves the real-space evolution of individual domain boundaries during a strain cycle, leaving open how strain-driven changes in local magnetic orientation are related to domain repopulation and domain-wall motion.
This distinction is particularly important in $\alpha$-MnTe because the weak out-of-plane moment is related to the in-plane N\'eel-vector angle through $M_z\propto\sin(3\phi_L)$, such that each sign of the out-of-plane moment represents three possible $120^{\circ}$-separated orientations of $\mathbf{L}$
\cite{
    Kriegner2017PRB,
    Gonzalez2023PRL,
    Kluczyk2024PRB}.

\begin{figure}[!htbp]
\centering
\includegraphics[width=0.99\textwidth]{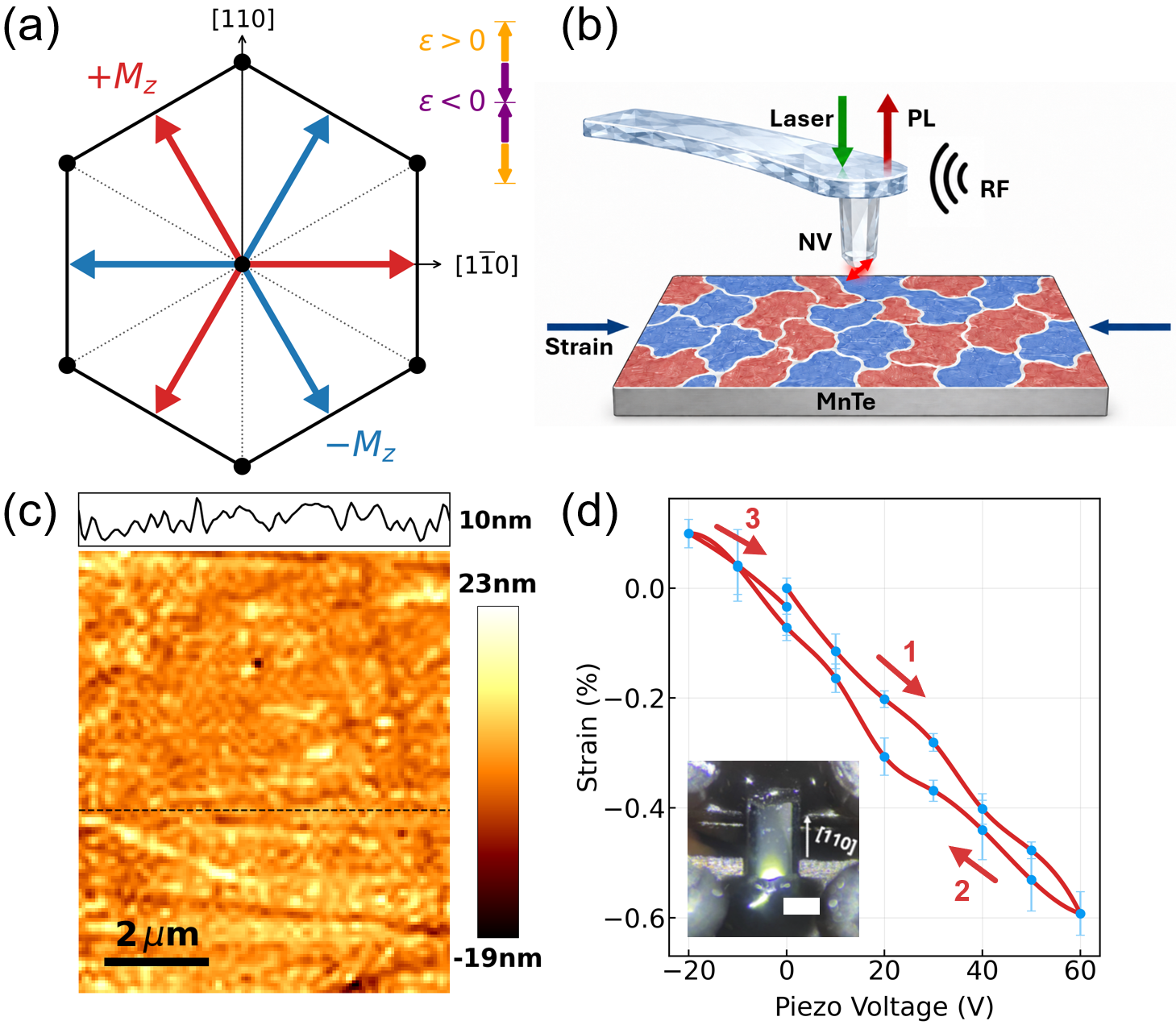}
\caption{\label{fig:1} 
(a) Schematic of $\alpha$-MnTe crystal structure. Mn atoms are shown as black circles and Te atoms are omitted for clarity. Red/blue arrows correspond to the six N\'eel vector easy axes, each associated with one of two out-of-plane (OOP) magnetic moments $\pm M_z$. Purple/orange arrows highlight compressive/tensile strain directions respectively along [110] axis.
(b) Schematic of scanning NV measurement. A diamond-tipped probe containing a single NV is scanned over a sample consisting of $\pm M_z$ regions (red/blue) associated with various N\'eel vector orientations. Excitation laser light is applied and NV PL is collected through a microscope objective above to optically readout the stray field strength.
(c) AFM image of region of interest used in all subsequent figures. (Inset) Line cut of AFM scan across horizontal dashed line. Scale bar is 2 $\mu$m.
(d) Calibration of applied voltage to uniaxial strain, determined from the relative change in the separation between surface particles tracked in large-area AFM scans. Numbered arrows indicate the strain-sweep direction. (Inset) Optical image of the sample mounted in the strain cell, with uniaxial stress applied along the [110] crystallographic direction, corresponding to a nearest-neighbor Mn-Mn bond direction. Scale bar of inset: 0.5 mm.
}
\end{figure}

Here, to address this debate, we integrate a piezo-driven uniaxial strain cell with scanning nitrogen-vacancy (NV) magnetometry, enabling direct nanoscale imaging of the stray magnetic field in bulk $\alpha$-MnTe under continuously tunable strain at room temperature [See Fig.~\ref{fig:1}(b)].
The uniaxial strain within the region of interest is quantified from atomic force microscopy (AFM) topography acquired with the same scanning NV probe.
Using this platform, we visualize the evolution of the magnetic-domain texture on the bulk $\alpha$ -MnTe surface at room temperature through a complete strain cycle from zero strain to maximum compression, small tension, and back to zero.
%
We find that increasing compressive strain promotes the coalescence of NV-resolved domains, enlarging the maximum connected-domain area while reducing the visible domain-wall density.
%
This evolution provides direct microscopic evidence that strain drives the near-surface magnetic texture toward a strain-selected domain configuration through domain-wall motion and domain repopulation.
During the unloading of the strain, however, the domain network follows a different trajectory and fragments into a metastable configuration, producing pronounced hysteresis in the maximum domain size and in the local stray-field distribution, despite comparatively weak hysteresis in the domain wall density.
%
%
Our observations therefore establish domain-boundary motion, coalescence, pinning, and metastability as key components of strain control in $\alpha$-MnTe. Although these measurements do not resolve the local in-plane N\'eel-vector angle and therefore cannot test continuous rotation directly, they demonstrate that a purely homogeneous rotation of an otherwise fixed domain texture is insufficient to describe the observed response. Together, these results establish scanning NV magnetometry under in-situ uniaxial strain as a platform for resolving and manipulating altermagnetic-domain dynamics at the nanoscale.

\section{Experimental results}

Single crystals of $\alpha$-MnTe were grown using a Te self-flux method~\cite{Liu2025arXiv}. Elemental Mn powder (99.95\%) and Te pieces (99.9999\%; both from Thermo Scientific Chemicals) were loaded into an alumina crucible in a Mn:Te molar ratio of $42:58$, with a total charge mass of $\sim$30~g. The crucible was sealed under vacuum in a quartz ampule, heated to $1020~^{\circ}\mathrm{C}$ over 10~h and held at this temperature for 12~h to obtain a homogeneous melt. The melt was then cooled to $800~^{\circ}\mathrm{C}$ at a rate of $0.6~^{\circ}\mathrm{C}/\mathrm{h}$, after which the excess Te flux was removed by centrifugation. Typical crystals had dimensions of approximately $10\times5\times5~\mathrm{mm}^{3}$.

$\alpha$-MnTe crystallizes in the hexagonal NiAs structure and hosts a collinear compensated antiferromagnetic order, in which Mn$^{2+}$ moments are aligned ferromagnetically within each basal-plane layer and antiferromagnetically between adjacent layers along the $c$ axis
\cite{
    Kriegner2017PRB,
    Gonzalez2023PRL, 
    Kluczyk2024PRB}.
Although the antiferromagnetic spin sublattices are nearly compensated, symmetry permits a weak net magnetic moment along the out-of-plane (OOP) direction.
The microscopic origin of this weak OOP  moment remains an active subject of discussion, particularly regarding the relative roles of spin canting, orbital magnetization, surface symmetry breaking, and higher-order spin-orbit-coupled interactions
\cite{mazinAltermagnetismMnTeOrigin2023a,mazinOriginGossamerFerromagnetism2024a,ChenYe2026Dominant}.
Similar OOP moment has been observed in altermagnet candidate h-FeS, and the OOP moment is found to be tunable under uniaxial strain directly associated with strain induced changes in AHE~\cite{Yao2026hFeS}.  
Within recent symmetry-based descriptions, the weak OOP moment is coupled to the in-plane N\'eel vector through a higher-order spin-orbit-coupled invariant, rather than through a conventional Dzyaloshinskii-Moriya-type spin-canting mechanism that would generate an effective bilinear coupling between $\mathbf{L}$ and $\mathbf{M}$
\cite{mazinOriginGossamerFerromagnetism2024a,Mcclarty2024,zhou2026imaging}.
As a result, the sign of the weak OOP moment is locked to the in-plane orientation of $\mathbf{L}$, but the measured sign of $M_z$ does not uniquely specify the full in-plane N\'eel-vector direction.

The six in-plane N\'eel-vector orientations of $\alpha$-MnTe form three crystallographically equivalent easy axes along the next-nearest-neighbor Mn--Mn bonds, each comprising a time-reversal pair, $\mathbf{L}$ and $-\mathbf{L}$, with opposite weak OOP moments (Fig.~\ref{fig:1}a)
\cite{Aoyama2024PRMaterials,Kluczyk2024PRB,Liu2025arXiv}.
Scanning NV magnetometry therefore directly resolves the sign and spatial distribution of the OOP stray fields, although each measured sign remains compatible with three distinct N\'eel-vector orientations.
This symmetry-imposed degeneracy prevents a unique assignment of the local N\'eel-vector direction from $B_z$ alone, but still permits robust analysis of domain boundaries, connectivity, and strain-driven reconfiguration.
Conversely, selecting a single in-plane N\'eel-vector axis does not uniquely determine the sign of $M_z$, because the two time-reversal-related states within that axis carry opposite weak OOP moments.
Thus, the previously reported single-domain state identified by orientation-sensitive measurements refers to the selection of one in-plane N\'eel axis, while positive- and negative-$M_z$ domains may still coexist within the corresponding time-reversal pair.

Figure~\ref{fig:1}b shows the experimental setup: a single NV center at the tip of a scanning probe senses the projection of the stray field $B_\parallel(x,y)=\vb{B}(x,y)\cdot\hat{\vb{n}}_\text{NV}$ on the NV axis $\hat{\vb{n}}_\text{NV}$ as a function of position above the sample and at a height of $\sim$\,40\,nm.
The OOP magnetic field $B_z(x,y)$ is calculated from $B_\parallel$ using a Fourier-space reconstruction that enforces the source-free condition $\nabla\cdot\mathbf{B}=0$ above the sample and the known NV spin axis.
The effective OOP magnetization $M_z(x,y)$ can then be calculated via a gradient-descent algorithm \cite{q312-kf83}.
However, because the $B_z(x,y)$ and $M_z(x,y)$ maps are qualitatively similar (see Supplementary Note 2), the analysis and conclusions in this work are based primarily on $B_z(x,y)$.
Accordingly, we use $B_z$ when referring to the measured stray field and $M_z$ when referring to the corresponding OOP magnetic moment.

As shown in Figure~\ref{fig:1}d, an $\alpha$-MnTe bulk crystal sample was mounted on a home-built strain cell \cite{hicksPiezoelectricbasedApparatusStrain2014a,malinowskiSuppressionSuperconductivityAnisotropic2020} capable of applying precise uniaxial stress along the NN Mn-Mn bond direction. 
The sample surface was mechanically polished, and AFM scan was taken at the location where the magnetometry was performed, showing a root-mean-square surface roughness of $h_\text{rms}=3.7$\,nm (Fig.~\ref{fig:1}c), confirming a smooth surface suitable for close-proximity NV magnetometry.
The actual strain applied on the sample is determined from the fractional change in distance between identifiable surface features tracked in large-area AFM scans (see Supplementary Note 1). As shown in Fig.~\ref{fig:1}d, the measured strain varies approximately linearly with the voltage applied to the strain cell, with a small hysteresis arising from the intrinsic ferroelectric response of the piezoelectric actuator \cite{hicksPiezoelectricbasedApparatusStrain2014a}.


\begin{figure}[!htbp]
\centering
\includegraphics[width=0.99\textwidth]{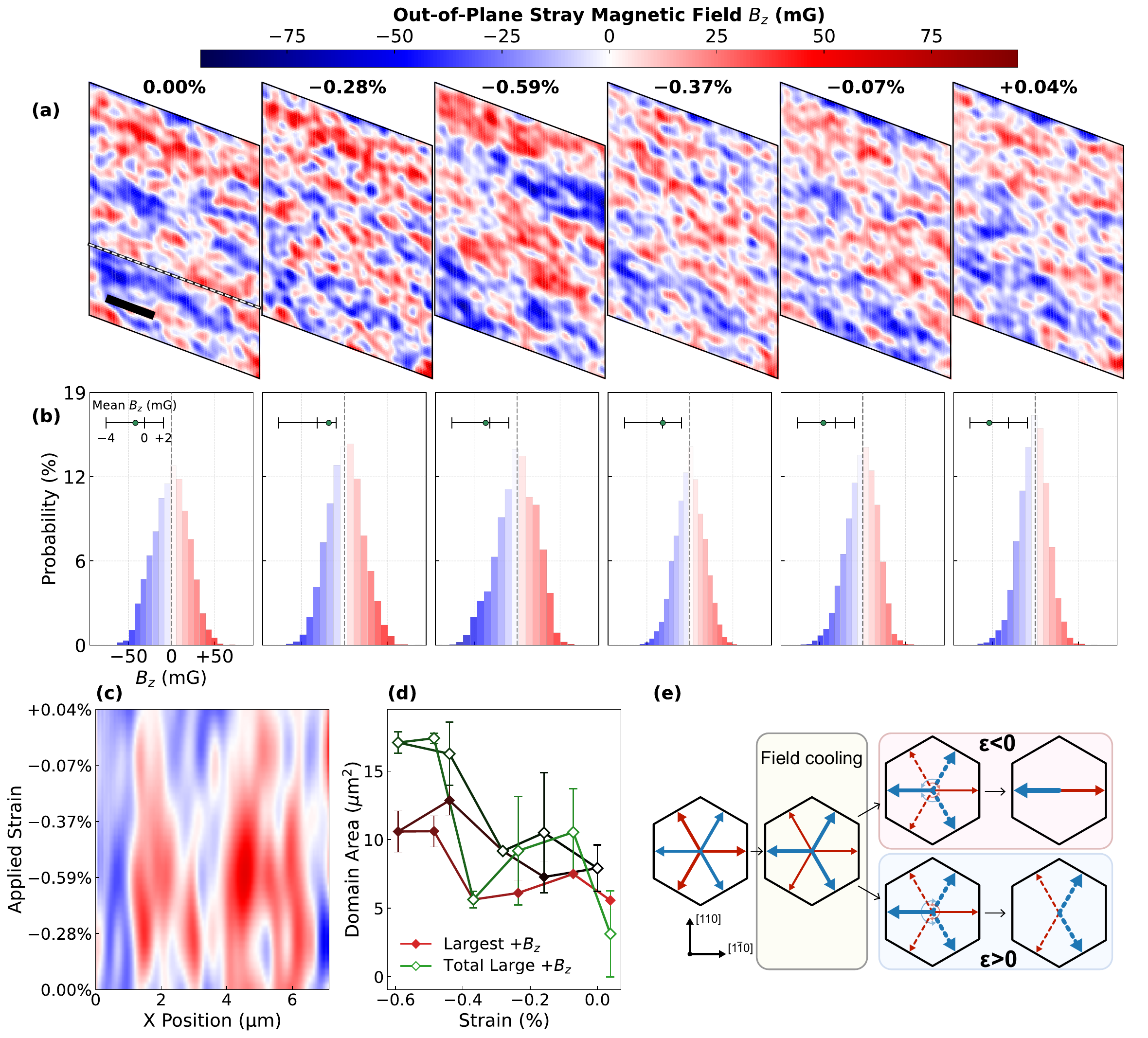}
\caption{\label{fig:2} 
(a) $B_z$ component extracted from NV magnetometry scans during a complete strain cycle, from 0\% strain to a maximum compressive strain of $\sim-0.6$\%, then back to $0.04$\% tensile strain. The dashed white line in the leftmost scan represents the location of line cuts presented in (c). Uniaxial stress is applied along the horizontal x direction. Scale bar: $2\,\mu\mathrm{m}$.
(b) Associated $B_z$ histograms corresponding to scans in (a). Approximately 10,000 data points were used to form each histogram. (Inset) Mean value of $B_z$.
(c) Interpolated linecut comparison of $B_z$ values at white/black dashed line in (a), shown as a function of strain.
(d) Maximum positive-domain area and total large-domain area as a function of strain, with the latter defined as the total positive area contained in domains larger than $5\,\mu\mathrm{m}^2$. To differentiate up- from down-sweeps, line colors begin black and transitions to red/green respectively following the steps within the strain cycle shown in Fig.~\ref{fig:1}d.
\textcolor{black}{
(e) Schematic of a possible strain-induced domain-population pathway. Relative N\'eel-vector domain populations are indicated qualitatively by number of arrows and color saturation. OOP field cooling initially biases the population toward the three $-B_z$ variants (blue), whereas compressive strain lowers the energy of the strain-favored in-plane axis pair along $\pm[1\overline{1}0]$. Redistribution among the allowed variants can then produce unequal populations of the two time-reversal-related states and increase the observed $+B_z$ sector (red). 
}
}
\end{figure}

Figure~\ref{fig:2} presents the spatial evolution of the magnetic domains across a complete strain cycle.
In this work, we focused primarily on compressive strain with minor tension because larger tensile strain could easily break the sample at room temperature.
After each compression cycle, a small negative voltage was applied to drive the strain cell slightly into the tensile regime, thereby compensating for piezoelectric hysteresis and completing the strain cycle.
The NV magnetometry maps in Figure~\ref{fig:2}a reveal a complex multidomain landscape established following the initial out-of-plane field-cooling procedure.
As compressive strain is swept along the $[110]$ axis and subsequently released, the local OOP stray-field texture undergoes substantial reorganization and does not return to its initial configuration.
The mean value of $B_z$ shifts toward more positive values under increasing compression and moves back toward more negative values as the compression is released (Figure~\ref{fig:2}b inset).
This response is particularly notable because an in-plane, time-reversal-even mechanical perturbation produces a pronounced reorganization of the out-of-plane stray-field texture. It implys an unusual cross-axis magnetomechanical response linking strain control of the in-plane N\'eel order to the NV-resolved $B_z$ signal.
Likewise, interpolated linecuts in Figure~\ref{fig:2}c show the expansion of positive-$B_z$ regions at large compressive strain and their recession or fragmentation on the release branch.
For comparison, additional room-temperature NV maps scanned at another region are shown in Supplementary Note 4, and scans acquired above the N\'eel temperature ($T_N=307$\,K) are shown in Supplementary Note 5, the latter confirming the absence of magnetic domains in the paramagnetic phase \cite{kriegnerMultiplestableAnisotropicMagnetoresistance2016}.


To quantify this reorganization, Figure~\ref{fig:2}d tracks the maximum positive-domain area and the total positive area contained in domains larger than $5\,\mu\mathrm{m}^2$.
Both quantities increase during the increasing compressive strain, demonstrating that compression promotes the formation of large connected positive-$B_z$ regions.
This behavior indicates that the positive projected OOP-moment sector does not grow primarily through the nucleation of many isolated small domains.
Instead, compressive strain drives boundary-mediated coalescence, in which neighboring regions with the compatible projected $M_z$ sign merge into larger connected domains.
%
%
The simultaneous increase of the largest connected positive domain and the total large-domain area shows that the upper tail of the positive-domain size distribution grows under compressive strain.
The texture nevertheless remains multi-domain over the strain range studied here, indicating that coalescence is likely limited by pinning, disorder, or thermal fluctuations on the surface rather than proceeding to complete single domain formation at low temperatures \cite{zhou2026imaging,Liu2025arXiv}.
The reverse strain sweep reveals that this coalescence is not simply reversed.
When the compressive strain is reduced from its maximum value, the largest connected positive domain decreases rapidly and became smaller than in the initial nominally zero-strain state.
The total large-domain area follows a similar hysteretic trend.
Thus, the reverse sweep appears to fragment, or depercolate, the strain-formed positive-domain network into a new metastable configuration, rather than restoring the original domain pattern.

A simple geometric picture (Fig.\,\ref{fig:2}e) illustrates one possible strain-induced population-transfer pathway.
OOP field cooling initially biases the projected texture toward the $-B_z$ sector.
While the initial OOP field-cooling procedure did not fully saturate the sample, an observable bias toward $-B_z$ is evident in Fig.~\ref{fig:2}, consistent with polarized-neutron diffraction measurements showing that c-axis field cooling induces a finite antiphase-domain imbalance in bulk $\alpha$-MnTe \cite{Liu2026PolarizedNeutron}.
%
Under our experimental geometry, compressive strain lowers the energy of the N\'eel-vector axis pair associated with the strain-favored direction, in this case along $\pm[1\overline{1}0]$.
Simultaneously, the domain-wall density is lower because fewer domain configurations are allowed than in the tensile-strain state, in which the remaining four N\'eel-vector domains selected by tensile strain can coexist, as demonstrated by neutron scattering measurements~\cite{Liu2025arXiv}.
Assuming that the three $-B_z$ N\'eel-vector variants are approximately equally populated, if local regions reorient toward the nearest member of this strain-favored axis pair, the two final time-reversal-related directions need not acquire equal population.
%
As the strain is swept toward tension, additional N\'eel-vector domain states become energetically accessible, which can increase the domain-wall density. Because the local N\'eel vector rotates continuously within each wall -- precisely where the cubic altermagnetic invariant $M_z\propto\sin{3\phi_L}$ vanishes by symmetry -- the growing wall area suppresses the effective OOP moment averaged over the imaged region.

Thus, an initially OOP-biased but in-plane-degenerate texture can be converted by compressive strain into an unequal population of the two time-reversal-related states within the selected in-plane axis pair.
This provides a possible microscopic pathway by which strain-induced axis selection, together with the memory of the field-cooled OOP imbalance, produces the observed growth of the positive-$B_z$ sector. 
%


\begin{figure}[!htbp]
\centering
\includegraphics[width=0.99\textwidth]{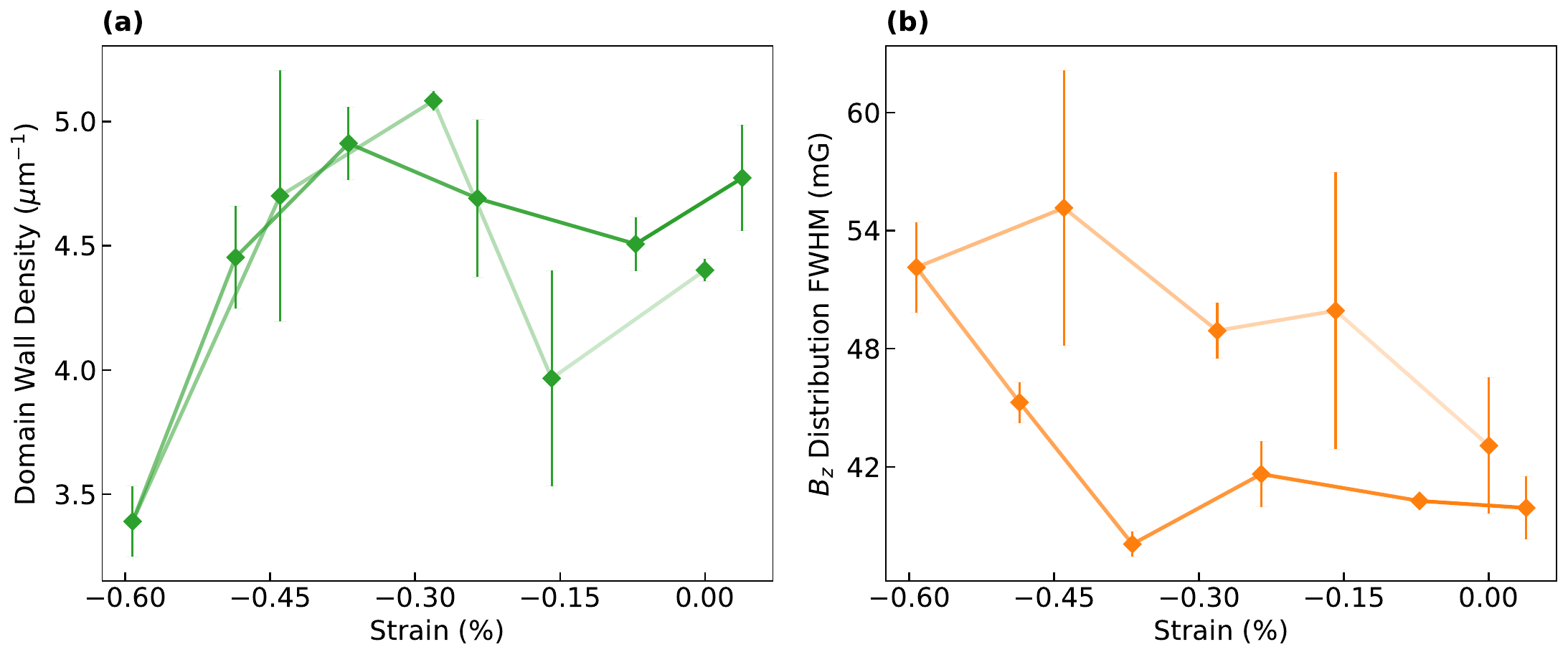}
\caption{\label{fig:3} 
(a) Average domain wall density and
(b) FWHM of $B_z$ distributions (see Fig.~\ref{fig:2}b for histograms), as functions of compressive strain along $[110]$.
Faded lines correspond to the first earlier measurements, with more solid lines representing later measurements.
}
\end{figure}

%


Having established strain-induced changes in the connectivity of the positive-$B_z$ network, we next compare this connectivity response with two complementary metrics: the visible domain-wall density and the full width at half-maximum (FWHM) of the $B_z$ distribution.
The domain-wall density plotted in Figure~\ref{fig:3}a was extracted from the $B_z$ domain masks and provides a measure of the visible boundary length per unit area (see Supplementary Note 3).
This quantity exhibits a weakly non-monotonic response at low-to-intermediate compressive strain, followed by a more pronounced decrease at the largest compression.
One possible interpretation is that, before net coalescence dominates, the evolving magnetoelastic anisotropy first mobilizes and reshapes the existing domain network, allowing boundaries to move, reconnect, and locally increase the total visible wall length.
%
%
At larger compressive strain, the concurrent growth of large connected positive-$B_z$ regions in Fig.~\ref{fig:2}d and reduction in visible domain-wall density in Fig.~\ref{fig:3}a support a crossover to domain coalescence and boundary elimination.
The weak hysteresis of this metric indicates that the total visible boundary length is governed mainly by the instantaneous applied external strain, rather than by the long-lived memory encoded in the domain-network connectivity.
Thus, the domain-wall-density trend is consistent with a crossover from strain-induced wall reconfiguration at low-to-intermediate compressive strain to net boundary elimination at larger compression, rather than with uniform coarsening throughout the full strain range.

The FWHM of the $B_z$ distribution extracted from Figure~\ref{fig:2}b captures the global spread of local OOP stray fields within the scan area (Figure~\ref{fig:3}b).
Because the magnetic stray field is measured at a finite NV--sample distance, the measured stray field of small or highly intermixed regions of opposite OOP moment partially cancel at NV height and produce reconstructed $B_z$ values closer to zero, leading to larger number of values near the center of the histogram.
During loading, compressive strain promotes the coalescence of smaller domains into larger connected domains.
This reduces local field cancellation and increases the fraction of the scanned area occupied by larger domains, thereby increasing the likelihood that the NV center detects larger absolute stray fields during the pixel-by-pixel scan.
Probability is therefore transferred from the central part of the histogram toward the sides, producing the broader $B_z$ distribution observed on the increasing-compression branch in Figure~\ref{fig:3}b.
During strain unloading, the strain-formed connected regions fragment or lose connectivity as the magnetoelastic driving force is reduced.
The resulting finer mixture of positive- and negative-$B_z$ regions increases local field cancellation and transfers probability back toward the center of the histogram, narrowing the $B_z$ distribution.
Because the unloading process does not reconstruct the original domain configuration, the return-branch FWHM can fall below its initial value.
Thus, the matching hysteresis of the largest-domain size and the $B_z$-distribution width supports a common origin in loading-induced coalescence and unloading-induced breakup of large connected domains.

Taken together, the metrics in Figure~\ref{fig:3} separate the strain response
of the total visible boundary length from the response of the stray-field amplitude distribution.
The domain-wall density exhibits comparatively weakly hysteretic, indicating that
the total amount of visible positive/negative boundary per unit area is not
the primary memory variable during the strain cycle.
By contrast, the largest-domain metrics in Figure~\ref{fig:2}d and the
$B_z$-distribution width in Figure~\ref{fig:3}b exhibit pronounced hysteresis.
This contrast shows that similar total visible boundary lengths can coexist with substantially different domain configurations.
The strain memory is carried primarily by the connectivity and organization of the domain network, rather than simply by the total amount of domain wall.
The strain cycle is therefore comparatively reversible in total visible
boundary length, but strongly hysteretic in the connectivity and organization
of the projected OOP-moment texture and in the resulting distribution of
stray-field amplitudes.

\section{Discussion and Outlook}
By combining scanning NV magnetometry with in-situ uniaxial-strain control at room temperature, we directly resolve the real-space evolution of the magnetic-domain texture at the surface of bulk $\alpha$-MnTe through a complete strain cycle.
Increasing compressive strain reorganizes and coalesces the domains, whereas unloading fragments the strain-formed connected regions into a metastable configuration.
The resulting hysteresis is most pronounced in metrics sensitive to domain connectivity, rather than in the total visible domain-wall density, identifying the organization of the domain network as a key carrier of strain-induced magnetic memory.
More broadly, the ability of an in-plane mechanical perturbation to reconfigure an OOP stray-field texture represents an unusual form of cross-axis control in a nearly compensated magnet.

This real-space evolution clarifies the relationship between strain-induced domain repopulation and the proposed continuous-rotation picture.
%
Our measurements establish that domain-boundary motion and reorganization of the projected $M_z$ domain network are essential components of the near-surface strain response. A homogeneous-rotation-only description without domain-network reorganization is therefore insufficient.
%
%
Continuous N\'eel-vector rotation and domain-population selection may coexist, but they correspond to distinct observables and should not be regarded as experimentally equivalent. Scanning NV magnetometry resolves changes in the projected OOP-moment texture but does not uniquely determine the local in-plane N\'eel-vector angle. Our measurements therefore neither confirm nor exclude continuous rotation within individual domains; rather, they show that the observed strain response necessarily includes substantial domain-boundary motion and history-dependent domain reorganization.
%

The observed hysteresis also provides a possible microscopic basis for the persistent strain- and history-dependent AHE in $\alpha$-MnTe
\cite{Liu2025arXiv,smolenski2025strain}.
Because the anomalous Hall conductivity depends on N\'eel-vector orientation and the relative populations of magnetic domains, strain-driven changes in domain orientations and populations can modify the spatially averaged transport response \cite{Gonzalez2023PRL,Mcclarty2024,Kluczyk2024PRB,Liu2025arXiv}.
As demonstrated in strain-dependent transport experiments, both the emergent AHE below 200 K and the unconventional scaling between anomalous Hall conductivity and longitudinal conductivity are observed only in the strain-stabilized single-domain state of $\alpha$-MnTe \cite{Liu2025arXiv,smolenski2025strain}.

The incomplete recovery of the domain texture further suggests that the Hall response may depend not only on the instantaneous strain, but also on the preceding strain trajectory.
However, scanning NV magnetometry probes the stray field associated with the weak OOP moment and does not uniquely determine either the local N\'eel-vector angle or the anomalous Hall conductivity.
Combining scanning magnetometry, orientation-sensitive imaging and transport measurements on the same strained device would therefore enable a direct correlation between local domain rearrangements and macroscopic electrical readout.
Time-resolved measurements and repeated strain cycles could further reveal domain-wall depinning, relaxation and training.
Nevertheless, both scanning NV magnetometry and magneto-optical measurements are predominantly sensitive to near-surface magnetic responses and may not directly represent the bulk domain configuration. 
In addition, the magnetic-domain size exhibits a pronounced thickness dependence \cite{zhou2026imaging}. 
Disorder may further influence the observed behavior by providing pinning sites that hinder domain-wall motion and domain reorganization. The same impurities/disorders have also been proposed to facilitate the AHE, possibly through modification of the charge gap \cite{Liu2025arXiv}, and even to induce local symmetry breakings in $\alpha$-MnTe~\cite{ren2026atomic}. 
Thin-film growth studies indicate that a high Te/Mn flux ratio and elevated growth temperature favor the formation of high-quality $\alpha$-MnTe phase \cite{shao2026epitaxial}. No AHE was reported for the crystals used in those studies~\cite{liebman2026strain,jost2025chiral}, suggesting that differences in growth conditions relative to those of our samples may have led to variations in crystalline quality, stoichiometry, and disorder, potentially contributing to the distinct magnetic and transport responses. 
A comprehensive understanding of the magnetism in bulk $\alpha$-MnTe under strain therefore requires disentangling near-surface effects, thickness-dependent domain formation, disorder-induced pinning, and growth-dependent sample properties~\cite{RevModPhys.82.1539,belashchenkoGiantStrainInducedSpin2025,osin2025extrinsic,zhao2026emergent,liuScalingAnomalousHall2011a,Mcclarty2024,Fernandes2024,Takahashi2025,hu2025catalog,Gonzalez2023PRL}. 

Beyond $\alpha$-MnTe, this capability provides a platform for mechanically programming compensated magnetic textures.
The observed history dependence could enable strain-written nonvolatile or multilevel memory, reconfigurable altermagnetic states and programmable spintronic networks controlled electrically through piezoelectric actuation.
More broadly, the ability to image compensated magnetic textures while continuously tuning their mechanical energy landscape establishes a route for understanding and ultimately engineering the nonequilibrium domain processes that govern strain-based altermagnetic memory, reconfigurable spintronics and mechanically programmable quantum materials.

\newcommand{\acknowtext}{%
The scanning NV microscopy was supported by the Center for Nanophase Materials Sciences, (CNMS), which is a US Department of Energy, Office of Science User Facility at Oak Ridge National Laboratory.
This research project was supported by the Laboratory Directed Research and Development Program of Oak Ridge National Laboratory, managed by UT-Battelle, LLC, for the U.S.~Department of Energy.
Part of the data analysis and sample preparation research were supported by the U.S. Department of Energy (DOE), Office of Science, Basic Energy Sciences, Materials Sciences and Engineering Division.
This manuscript has been authored by UT-Battelle, LLC, under contract DE-AC05-00OR22725 with the US Department of Energy (DOE).
Single-crystal growth at Rice University was supported by the U.S. Department of Energy, Office of Basic Energy Sciences, under Award No. DE-SC0012311 and DE-SC0026179 (P.D.). Part of the materials characterization efforts at Rice is supported by
the Robert A. Welch Foundation Grant No. C-1839 (P.D.).
The authors thank J.-H. Chu for fruitful discussion. 
The US government retains and the publisher, by accepting the article for publication, acknowledges that the US government retains a nonexclusive, paid-up, irrevocable, worldwide license to publish or reproduce the published form of this manuscript, or allow others to do so, for US government purposes.
DOE will provide public access to these results of federally sponsored research in accordance with the DOE Public Access Plan (http://energy.gov/downloads/doe-public-access-plan).
}
\paragraph{Acknowledgments:}
\acknowtext

\newpage
\bibliography{references.bib}

@article{ren2026atomic,
  title={Atomic-Scale Observation of Symmetry Breaking in Altermagnetic MnTe},
  author={Ren, Guodong and DeStefano, Jonathan M and Zhang, Xiao-Wei and Thind, Arashdeep S and Giridharagopal, Rajiv and Castellanos-Reyes, Jose Angel and Zeiger, Paul M and Kamm, Noah and Liu, Zhaoyu and Xie, Yaofeng and others},
  journal={arXiv preprint arXiv:2605.27543},
  year={2026}
}

@article{Yao2026hFeS,
author = {Yao, Weiliang and Ye, Feng and Morgan, Zachary J. and Abernathy, Douglas L. and Liu, Ruixian and Xu, Sijie and Gao, Yuxiang and Allen, Kevin and Fang, Yuan and Morosan, Emilia and Si, Qimiao and Dai, Pengcheng},
title = {Uniaxial Strain Tuned Magnetism of the Altermagnet Candidate H-FeS},
journal = {Advanced Materials},
volume = {n/a},
number = {n/a},
pages = {e74398},
doi = {https://doi.org/10.1002/adma.74398},
url = {https://advanced.onlinelibrary.wiley.com/doi/abs/10.1002/adma.74398},
}

@article{Liu2026PolarizedNeutron,
  title = {Observation of Altermagnetic Order Switching in Bulk MnTe by Polarized Neutron Diffraction},
  author = {Liu, Zheyuan and Asai, Shinichiro and Takahashi, Shingo and Saito, Hiraku and Nakajima, Taro and Masuda, Takatsugu},
  journal = {Phys. Rev. Lett.},
  volume = {136},
  issue = {25},
  pages = {256704},
  numpages = {8},
  year = {2026},
  month = {Jun},
  publisher = {American Physical Society},
  doi = {10.1103/yklc-9n6t},
  url = {https://link.aps.org/doi/10.1103/yklc-9n6t}
}

@article{Smejkal2022PRX_beyond,
  title = {Beyond Conventional Ferromagnetism and Antiferromagnetism: A Phase with Nonrelativistic Spin and Crystal Rotation Symmetry},
  author = {{\v{S}}mejkal, Libor and Sinova, Jairo and Jungwirth, Tomas},
  journal = {Physical Review X},
  volume = {12},
  number = {3},
  pages = {031042},
  year = {2022},
  publisher = {American Physical Society},
  doi = {10.1103/PhysRevX.12.031042}
}

@article{Smejkal2022PRX_emerging,
  title = {Emerging Research Landscape of Altermagnetism},
  author = {\ifmmode \check{S}\else \v{S}\fi{}mejkal, Libor and Sinova, Jairo and Jungwirth, Tomas},
  journal = {Physical Review X},
  volume = {12},
  issue = {4},
  pages = {040501},
  numpages = {27},
  year = {2022},
  month = {Dec},
  publisher = {American Physical Society},
  doi = {10.1103/PhysRevX.12.040501},
  url = {https://link.aps.org/doi/10.1103/PhysRevX.12.040501}
}

@article{Smejkal2022NatRev,
  title = {Anomalous Hall antiferromagnets},
  author = {{\v{S}}mejkal, Libor and MacDonald, A. H. and Sinova, Jairo and Nakatsuji, S. and Jungwirth, Tomas},
  journal = {Nature Reviews Materials},
  volume = {7},
  pages = {482--496},
  year = {2022},
  publisher = {Nature Publishing Group},
  doi = {10.1038/s41578-022-00430-3}
}

@article{Krempasky2024Nature,
  title = {Altermagnetic lifting of Kramers spin degeneracy},
  author = {Krempask{\'y}, J. and {\v{S}}mejkal, L. and D'Souza, S. W. and Hajlaoui, M. and Springholz, G. and Uhl{\'\i}{\v{r}}ov{\'a}, K. and Alarab, F. and Constantinou, P. C. and Strocov, V. and Usanov, D. and others},
  journal = {Nature},
  volume = {626},
  pages = {517--522},
  year = {2024},
  publisher = {Nature Publishing Group},
  doi = {10.1038/s41586-023-06907-7}
}

@article{Gonzalez2023PRL,
  title = {Spontaneous Anomalous Hall Effect Arising from an Unconventional Compensated Magnetic Phase in a Semiconductor},
  author = {Gonz{\'a}lez-Betancourt, R. D. and Zub{\'a}{\v{c}}, J. and Gonz{\'a}lez-Hern{\'a}ndez, R. and Geishendorf, K. and {\v{S}}ob{\'a}{\v{n}}, Z. and Springholz, G. and Olejn{\'\i}k, K. and {\v{S}}mejkal, L. and Sinova, J. and Jungwirth, T. and others},
  journal = {Physical Review Letters},
  volume = {130},
  number = {3},
  pages = {036702},
  year = {2023},
  publisher = {American Physical Society},
  doi = {10.1103/PhysRevLett.130.036702}
}

@article{Liu2025arXiv,
  title = {Strain-tunable anomalous Hall effect in hexagonal MnTe},
  author = {Liu, Zhaoyu and Xu, Sijie and DeStefano, Jonathan M. and Rosenberg, Elliott and Zhang, Tingjun and Li, Jinyulin and Stone, Matthew B. and Ye, Feng and Tian, Wei and Edwards, Sarah and others},
  year = {2025},
  journal = {arXiv:2509.19582},
  archivePrefix = {arXiv},
}

@article{Aoyama2024PRMaterials,
  title = {Piezomagnetic properties in altermagnetic {MnTe}},
  author = {Aoyama, T. and Ohgushi, K.},
  journal = {Physical Review Materials},
  volume = {8},
  number = {4},
  pages = {L041402},
  year = {2024},
  publisher = {American Physical Society},
  doi = {10.1103/PhysRevMaterials.8.L041402}
}

@article{Kriegner2017PRB,
  title = {Magnetic anisotropy in antiferromagnetic hexagonal MnTe},
  author = {Kriegner, D. and Reichlov{\'a}, H. and Grenzer, J. and Schmidt, W. and Ressouche, E. and Godinho, J. and Wagner, T. and Martin, S. Y. and Shick, A. B. and Volobuev, V. V. and others},
  journal = {Physical Review B},
  volume = {96},
  number = {21},
  pages = {214418},
  year = {2017},
  publisher = {American Physical Society},
  doi = {10.1103/PhysRevB.96.214418}
}

@article{Kluczyk2024PRB,
  title = {Coexistence of anomalous {Hall} effect and weak magnetization in a nominally collinear antiferromagnet {MnTe}},
  author = {Kluczyk, K. P. and Gas, K. and Grzybowski, M. J. and Skupinski, P. and Borysiewicz, M. A. and Fas, T. and Suffczynski, J. and Domagala, J. Z. and Grasza, K. and Mycielski, A. and others},
  journal = {Physical Review B},
  volume = {110},
  number = {15},
  pages = {155201},
  year = {2024},
  publisher = {American Physical Society},
  doi = {10.1103/PhysRevB.110.155201}
}

@article{hu2025catalog,
  title={Catalog of {$C$}-paired spin-momentum locking in antiferromagnetic systems},
  author={Hu, Mengli and Cheng, Xingkai and Huang, Zhenqiao and Liu, Junwei},
  journal={Physical Review X},
  volume={15},
  number={2},
  pages={021083},
  year={2025},
  publisher={APS},
  url = {https://journals.aps.org/prx/abstract/10.1103/PhysRevX.15.021083}
}

@article{smolenski2025strain,
      title={Strain-tunability of the multipolar {Berry} curvature in altermagnet {MnTe}}, 
      author={Shane Smolenski and Ning Mao and Dechen Zhang and Yucheng Guo and A. K. M. Ashiquzzaman Shawon and Mingyu Xu and Eoghan Downey and Trisha Musall and Ming Yi and Weiwei Xie and Chris Jozwiak and Aaron Bostwick and Nobumichi Tamura and Eli Rotenberg and Lu Li and Kai Sun and Yang Zhang and Na Hyun Jo},
      year={2025},
      journal={arXiv:2509.21481},
      url={https://arxiv.org/abs/2509.21481}, 
}

@article{liebman2026strain,
      title={Strain continuously rotates the N\'eel vector in altermagnetic {MnTe}}, 
      author={Alex Liebman-Peláez and Jon Kruppe and Resham Babu Regmi and Nirmal J. Ghimire and Yue Sun and Igor I. Mazin and Hilary M. L. Noad and James Analytis and Veronika Sunko and Joseph Orenstein},
      year={2026},
      journal={arXiv:2604.07653},
      url= {https://arxiv.org/abs/2604.07653} 
}

@article{osin2025extrinsic,
  title={Extrinsic anomalous {Hall} effect in altermagnets},
  author={Osin, A and Levchenko, A and Khodas, M},
  journal={arXiv:2511.03151},
  year={2025},
  url = {https://arxiv.org/abs/2511.03151}
}

@article{RevModPhys.90.015005,
  title = {Antiferromagnetic spintronics},
  author = {Baltz, V. and Manchon, A. and Tsoi, M. and Moriyama, T. and Ono, T. and Tserkovnyak, Y.},
  journal = {Rev. Mod. Phys.},
  volume = {90},
  issue = {1},
  pages = {015005},
  numpages = {57},
  year = {2018},
  month = {Feb},
  publisher = {American Physical Society},
  doi = {10.1103/RevModPhys.90.015005},
  url = {https://link.aps.org/doi/10.1103/RevModPhys.90.015005}
}

@article{Baral2023adfm,
author = {Baral, Raju and Abeykoon, A.M. Milinda and Campbell, Branton J. and Frandsen, Benjamin A.},
title = {Giant Spontaneous Magnetostriction in {MnTe} Driven by a Novel Magnetostructural Coupling Mechanism},
journal = {Advanced Functional Materials},
volume = {33},
number = {46},
pages = {2305247},
doi = {https://doi.org/10.1002/adfm.202305247},
year = {2023}
}

@article{RevModPhys.82.1539,
  title = {Anomalous {Hall} effect},
  author = {Nagaosa, Naoto and Sinova, Jairo and Onoda, Shigeki and MacDonald, A. H. and Ong, N. P.},
  journal = {Rev. Mod. Phys.},
  volume = {82},
  issue = {2},
  pages = {1539--1592},
  numpages = {0},
  year = {2010},
  month = {May},
  publisher = {American Physical Society},
  doi = {10.1103/RevModPhys.82.1539},
  url = {https://link.aps.org/doi/10.1103/RevModPhys.82.1539}
}

@article{belashchenkoGiantStrainInducedSpin2025,
  title = {Giant Strain-Induced Spin Splitting Effect in {MnTe}, a $g$-Wave Altermagnetic Semiconductor},
  author = {Belashchenko, K. D.},
  year = {2025},
  journaltitle = {Physical Review Letters},
  journal = {Phys. Rev. Lett.},
  volume = {134},
  number = {8},
  pages = {086701},
  publisher = {American Physical Society},
  doi = {10.1103/PhysRevLett.134.086701},
  url = {https://link.aps.org/doi/10.1103/PhysRevLett.134.086701},
  urldate = {2025-05-07}
}

@article{malinowskiSuppressionSuperconductivityAnisotropic2020,
  title = {Suppression of Superconductivity by Anisotropic Strain near a Nematic Quantum Critical Point},
  author = {Malinowski, Paul and Jiang, Qianni and Sanchez, Joshua J. and Mutch, Joshua and Liu, Zhaoyu and Went, Preston and Liu, Jian and Ryan, Philip J. and Kim, Jong Woo and Chu, Jiun Haw},
  year = {2020},
  journal = {Nature Physics},
  volume = {16},
  number = {12},
  pages = {1189--1193},
  publisher = {Springer US},
  issn = {17452481},
  doi = {10.1038/s41567-020-0983-9}
}

@article{hicksPiezoelectricbasedApparatusStrain2014a,
  title = {Piezoelectric-Based Apparatus for Strain Tuning},
  author = {Hicks, Clifford W. and Barber, Mark E. and Edkins, Stephen D. and Brodsky, Daniel O. and Mackenzie, Andrew P.},
  year = {2014},
  month = jun,
  journal = {Review of Scientific Instruments},
  volume = {85},
  number = {6},
  pages = {065003},
  publisher = {American Institute of Physics},
  issn = {0034-6748},
  doi = {10.1063/1.4881611},
  urldate = {2022-11-04}
}

@article{shao2026epitaxial,
  title={Epitaxial Growth and Anomalous Hall Effect in High-Quality Altermagnetic $\alpha$-MnTe Thin Films},
  author={Shao, Tian-Hao and Dai, Xingze and Hu, Wenyu and Zhu, Ming-Yuan and He, Yuanqiang and Yang, Lin-He and Liu, Jingjing and Yang, Meng and Liu, Xiang-Rui and Shi, Jing-Jing and others},
  journal={arXiv preprint arXiv:2602.11645},
  year={2026},
  url = {https://arxiv.org/abs/2602.11645}
}

@article{jost2025chiral,
  title={Chiral altermagnon in MnTe},
  author={Jost, Daniel and Regmi, Resham B and Lomeli, Eder G and Sahel-Schackis, Samuel and Scheufele, Monika and Neuhaus, Marcel and Nickel, Rachel and Yakhou, Flora and Kummer, Kurt and Brookes, Nicholas and others},
  journal={arXiv:2501.17380},
  year={2025},
  url = {https://arxiv.org/pdf/2501.17380}
}

@article{kriegnerMultiplestableAnisotropicMagnetoresistance2016,
  title = {Multiple-Stable Anisotropic Magnetoresistance Memory in Antiferromagnetic {{MnTe}}},
  author = {Kriegner, D. and Výborný, K. and Olejník, K. and Reichlová, H. and Novák, V. and Marti, X. and Gazquez, J. and Saidl, V. and Němec, P. and Volobuev, V. V. and Springholz, G. and Holý, V. and Jungwirth, T.},
  year = {2016},
  journaltitle = {Nature Communications},
  journal = {Nat Commun},
  volume = {7},
  number = {1},
  pages = {11623},
  publisher = {Nature Publishing Group},
  issn = {2041-1723},
  doi = {10.1038/ncomms11623},
  url = {https://www.nature.com/articles/ncomms11623},
  urldate = {2025-07-22},
  langid = {english}
}

@article{mazinAltermagnetismMnTeOrigin2023a,
  title = {Altermagnetism in {{MnTe}}: {{Origin}}, Predicted Manifestations, and Routes to Detwinning},
  shorttitle = {Altermagnetism in {{MnTe}}},
  author = {Mazin, I. I.},
  date = {2023},
  journaltitle = {Physical Review B},
  journal = {Phys. Rev. B},
  volume = {107},
  number = {10},
  pages = {L100418},
  publisher = {American Physical Society},
  doi = {10.1103/PhysRevB.107.L100418},
  url = {https://link.aps.org/doi/10.1103/PhysRevB.107.L100418}
}

@article{mazinOriginGossamerFerromagnetism2024a,
  title = {Origin of the Gossamer Ferromagnetism in {{MnTe}}},
  author = {Mazin, I. I. and Belashchenko, K. D.},
  date = {2024},
  journaltitle = {Physical Review B},
  journal = {Phys. Rev. B},
  volume = {110},
  number = {21},
  pages = {214436},
  issn = {2469-9950, 2469-9969},
  doi = {10.1103/PhysRevB.110.214436},
  url = {https://link.aps.org/doi/10.1103/PhysRevB.110.214436},
  urldate = {2025-08-13},
  langid = {english}
}

@article{songAltermagnetsNewClass2025,
  title = {Altermagnets as a New Class of Functional Materials},
  author = {Song, Cheng and Bai, Hua and Zhou, Zhiyuan and Han, Lei and Reichlova, Helena and Dil, J. Hugo and Liu, Junwei and Chen, Xianzhe and Pan, Feng},
  date = {2025-02-14},
  journaltitle = {Nature Reviews Materials},
  journal = {Nat Rev Mater},
  pages = {1--13},
  publisher = {Nature Publishing Group},
  issn = {2058-8437},
  doi = {10.1038/s41578-025-00779-1},
  url = {https://www.nature.com/articles/s41578-025-00779-1},
  urldate = {2025-05-11},
  langid = {english}
}

@article{liuScalingAnomalousHall2011a,
  title = {Scaling of the Anomalous {{Hall}} Effect in the Insulating Regime},
  author = {Liu, Xiong-Jun},
  year = {2011},
  journal = {Physical Review B},
  shortjournal = {Phys. Rev. B},
  volume = {84},
  number = {16},
  pages = {165304},
  doi = {10.1103/PhysRevB.84.165304}
}

@article{Takahashi2025,
  title = {Elasto-{Hall} conductivity and the anomalous {Hall} effect in altermagnets},
  author = {Takahashi, Keigo and Steward, Charles R. W. and Ogata, Masao and Fernandes, Rafael M. and Schmalian, J\"org},
  journal = {Phys. Rev. B},
  volume = {111},
  issue = {18},
  pages = {184408},
  numpages = {18},
  year = {2025},
  month = {May},
  publisher = {American Physical Society},
  doi = {10.1103/PhysRevB.111.184408},
  url = {https://link.aps.org/doi/10.1103/PhysRevB.111.184408}
}

@article{zhao2026emergent,
  title={Emergent anomalous Hall effect from surface states in altermagnetic MnTe thin films},
  author={Zhao, Yufei and Mandal, Saswata and Liu, Chao-Xing and Yan, Binghai},
  journal={Physical Review B},
  volume={114},
  number={6},
  pages={065304},
  year={2026},
  publisher={APS}
}

@article{jungwirth2026symmetry,
  title={Symmetry, microscopy and spectroscopy signatures of altermagnetism},
  author={Jungwirth, Tomas and Sinova, Jairo and Fernandes, Rafael M and Liu, Qihang and Watanabe, Hikaru and Murakami, Shuichi and Nakatsuji, Satoru and {\v{S}}mejkal, Libor},
  journal={Nature},
  volume={649},
  number={8098},
  pages={837--847},
  year={2026},
  publisher={Nature Publishing Group UK London},
  url = {https://www.nature.com/articles/s41586-025-09883-2#citeas}
}

@article{Mcclarty2024,
  title={{Landau theory of altermagnetism}},
  author={McClarty, Paul A and Rau, Jeffrey G},
  journal={Physical Review Letters},
  volume={132},
  number={17},
  pages={176702},
  year={2024},
  publisher={APS},
  doi={https://doi.org/10.1103/PhysRevLett.132.176702},
}

@article{Fernandes2024,
  title = {Topological transition from nodal to nodeless {Zeeman} splitting in altermagnets},
  author = {Fernandes, Rafael M. and de Carvalho, Vanuildo S. and Birol, Turan and Pereira, Rodrigo G.},
  journal = {Phys. Rev. B},
  volume = {109},
  issue = {2},
  pages = {024404},
  numpages = {19},
  year = {2024},
  month = {Jan},
  publisher = {American Physical Society},
  doi = {10.1103/PhysRevB.109.024404},
  url = {https://link.aps.org/doi/10.1103/PhysRevB.109.024404}
}

@article{q312-kf83,
  title = {Universal reconstruction of complex magnetic profiles with minimal prior assumptions},
  author = {Yao, Changyu and Yu, Yue and Shi, Yinyao and Jung, Ji-In and V\'aci, Zolt\'an and Wang, Yizhou and Liu, Zhongyuan and Zhang, Chuanwei and Tikoo-Schantz, Sonia and Zu, Chong},
  journal = {Phys. Rev. Appl.},
  volume = {24},
  issue = {6},
  pages = {064020},
  numpages = {12},
  year = {2025},
  month = {Dec},
  publisher = {American Physical Society},
  doi = {10.1103/q312-kf83},
  url = {https://link.aps.org/doi/10.1103/q312-kf83}
}

@article{zhou2026imaging,
      title={Imaging Surface Magnetization in Altermagnetic MnTe Films},
  author={Zhou, Ling-Jie and Li, Senlei and Yan, Zi-Jie and Zhao, Yufei and Rong, Hongtao and Xiong, Zelong and Zhao, Yiran and Xiao, Pu and Lai, Lok Kan and Bae, Hyeonhu and others},
  journal={arXiv preprint arXiv:2605.25241},
  year={2026} 
}

@article{ChenYe2026Dominant,
  title = {Dominant orbital magnetization in the prototypical altermagnet MnTe},
  author = {Chen Ye, Chao and Tenzin, Karma and S\l{}awi\ifmmode \acute{n}\else \'{n}\fi{}ska, Jagoda and Autieri, Carmine},
  journal = {Phys. Rev. B},
  volume = {113},
  issue = {1},
  pages = {014413},
  numpages = {9},
  year = {2026},
  month = {Jan},
  publisher = {American Physical Society},
  doi = {10.1103/g32j-hnvz},
  url = {https://link.aps.org/doi/10.1103/g32j-hnvz}
}

\end{document}